Original Paper

# Lamotrigine Therapy for Bipolar Depression: Analysis of Self-Reported Patient Data

Antoine Nzeyimana[1,2]; Kate EA Saunders[3,4]; John R Geddes[3,4]; Patrick E McSharry[2,5,6,7]

[1]Department of Geography, University of Oregon, Eugene, OR, United States
[2]Carnegie Mellon University Africa, Kigali, Rwanda
[3]University Department of Psychiatry, Warneford Hospital, Oxford, United Kingdom
[4]Oxford Health NHS Foundation Trust, Warneford Hospital, Oxford, United Kingdom
[5]African Center of Excellence in Data Science, University of Rwanda, Kigali, Rwanda
[6]Oxford-Man Institute of Quantitative Finance, University of Oxford, Oxford, United Kingdom
[7]Oxford Internet Institute, University of Oxford, Oxford, United Kingdom

**Corresponding Author:**
Antoine Nzeyimana
Department of Geography
University of Oregon
1251 University of Oregon
Eugene, OR, 97403-1251
United States
Phone: 1 541 346 0785
Email: anzeyima@uoregon.edu

## Abstract

**Background:** Depression in people with bipolar disorder is a major cause of long-term disability, possibly leading to early mortality and currently, limited safe and effective therapies exist. Although existing monotherapies such as quetiapine have limited proven efficacy and practical tolerability, treatment combinations may lead to improved outcomes. Lamotrigine is an anticonvulsant currently licensed for the prevention of depressive relapses in individuals with bipolar disorder. A double-blinded randomized placebo-controlled trial (comparative evaluation of Quetiapine-Lamotrigine [CEQUEL] study) was conducted to evaluate the efficacy of lamotrigine plus quetiapine versus quetiapine monotherapy in patients with bipolar type I or type II disorders.

**Objective:** Because the original CEQUEL study found significant depressive symptom improvements, the objective of this study was to reanalyze CEQUEL data and determine an unbiased classification accuracy for active lamotrigine versus placebo. We also wanted to establish the time it took for the drug to provide statistically significant outcomes.

**Methods:** Between October 21, 2008 and April 27, 2012, 202 participants from 27 sites in United Kingdom were randomly assigned to two treatments; 101: lamotrigine, 101: placebo. The primary variable used for estimating depressive symptoms was based on the Quick Inventory of Depressive Symptomatology—self report version 16 (QIDS-SR16). The original CEQUEL study findings were confirmed by performing *t* test and linear regression. Multiple features were computed from the QIDS-SR16 time series; different linear and nonlinear binary classifiers were trained to distinguish between the two groups. Various feature-selection techniques were used to select a feature set with the greatest explanatory power; a 10-fold cross-validation was used.

**Results:** From weeks 10 to 14, the mean difference in QIDS-SR16 ratings between the groups was –1.6317 ($P$=.09; sample size=81, 77; 95% CI –0.2403 to 3.5036). From weeks 48 to 52, the mean difference was –2.0032 ($P$=.09; sample size=54, 48; 95% CI –0.3433 to 4.3497). The coefficient of variation ($\sigma/\mu$) and detrended fluctuation analysis (DFA) exponent alpha had the greatest explanatory power. The out-of-sample classification accuracy for the 138 participants who reported more than 10 times after week 12 was 62%. A consistent classification accuracy higher than the no-information benchmark was obtained in week 44.

**Conclusions:** Adding lamotrigine to quetiapine treatment decreased depressive symptoms in patients with bipolar disorder. Our classification model suggested that lamotrigine increased the coefficient of variation in the QIDS-SR16 scores. The lamotrigine group also tended to have a lower DFA exponent, implying a substantial temporal instability in the time series. The performance of the model over time suggested that a trial of at least 44 weeks was required to achieve consistent results. The selected model confirmed the original CEQUEL study findings and helped in understanding the temporal dynamics of bipolar depression during treatment.

**Trial Registration:** EudraCT Number 2007-004513-33; https://www.clinicaltrialsregister.eu/ctr-search/trial/2007-004513-33/GB (Archived by WebCite at http://www.webcitation.org/73sNaI29O).

*(JMIR Ment Health 2018;5(4):e63)* doi: [10.2196/mental.9026](10.2196/mental.9026)

## Introduction

Bipolar disorder, a psychiatric condition characterized by repeated elevated mood (mania) and low mood (depression) states [1], has been ranked the sixth cause of disability worldwide, affecting nearly 1% of the adult population [2,3]. People with bipolar disorder [4] spend up to a third of their lives depressed, and it is these depressive symptoms that result in long-term disability and early mortality.

There is, however, no consensus on the effectiveness or safety of antidepressant drugs, such as fluoxetine, for bipolar depression [5]. Lamotrigine, a sodium channel inhibitor, is an anticonvulsant that is also being used to treat bipolar depression [6]; however, its action mechanism for bipolar disorder remains unclear [7]. Although Lamotrigine has been licensed for the prevention of bipolar disorder depressive relapses, its acute effects are less well known. A pooled analysis of all randomized data has revealed that although there appears to be a modest treatment effect in the acute phases of bipolar depression, this has not been observed in individual trials, which may have been because of the short treatment durations in these trials (8 weeks). Because lamotrigine requires a 6-week titration period, the majority of trials therefore only assessed 2 weeks of treatment at the therapeutic dose. There is still no consensus on the proven efficacy and practical tolerability of current monotherapies for bipolar depression such as quetiapine; however, it has been suggested that treatment combinations could lead to improved outcomes.

The comparative evaluation of quetiapine plus lamotrigine versus quetiapine monotherapy (CEQUEL) trial was a double-blind randomized placebo-controlled parallel group trial comparing a lamotrigine plus quetiapine treatment and a quetiapine monotherapy treatment in patients diagnosed with a bipolar I or II disorder (EudraCT Number: 2007-004513-33) [8]. A minimum level of depressive symptoms was not required for entry to either the run-in or randomized phases of the trial because the relevant criterion was the clinical judgment that new pharmacological treatments were required for depressive episodes. The randomization procedure used an adaptive minimization algorithm that was balanced for center, age, sex, bipolar disorder I or II, baseline depression severity, quetiapine dose, concurrent medication, pretrial use of quetiapine, pretrial use of lamotrigine, and mood episodes in past year (<4 or ≥4). The primary outcome measure for the trial was the presence of depressive symptoms at 12 weeks as self-reported by the subjects using the Quick Inventory of Depressive Symptomatology—self report version 16 (QIDS-SR16) [9].

Symptoms of mania were also assessed using the Altman self-rating scale (ASRM) [10]. The resulting scores ranged from 0-27 for the QIDS-SR16 and 0-25 for ASRM. The subjects were prompted weekly by text or email to report their mood symptoms using the True Colours platform, which is described in more detail in a study [11].

The original analysis reported significant depressive symptom improvements for the lamotrigine subjects compared with the placebo subjects. In this paper, the data collected in the trial were reanalyzed using machine learning approaches with the main objective being the identification of the most appropriate binary classifier to distinguish between the lamotrigine and the placebo effects. In addition to replicating the findings obtained from the original statistical analysis, we also wanted to determine the time it took for the drug to provide statistically significant outcomes to provide some guidance with respect to the minimum amount of time required to undertake trials that aim to establish the treatment or drug efficacy.

To assess the differences between patients taking lamotrigine and those taking the placebo, a binary classification approach was used to identify the relevant features to be extracted from the time series. This approach considered the different characteristics in the collected data and sought to classify the participants into 2 distinct groups based on the observed features; in this case, the group taking lamotrigine and the group taking the placebo. The features were determined based on the different statistical metrics computed from the data, and the analysis determined which features would facilitate the classification; for example, it was expected that the mean QIDS-SR16 scores would differ between the 2 groups. For every feature, a kernel based density estimation was used to examine its probability distribution between the 2 groups and to test whether the 2 distributions were different; the bigger the differences, the greater was the explanatory power of the feature. For the same reason, the performance of a classifier for each individual feature was also examined. Finally, the features and classifier with the best cross-validated accuracies were identified.

## Methods

### Data

Data were collected from 202 participants over a period of 52 weeks; 149 with bipolar type I disorder and 53 with bipolar type II disorder; of which 90 were male and 112 were female. Figure 1 shows the number of responses received each week over the 52 week data collection period.

**Figure 1.** Subject compliance: number of subjects reporting in a particular week.

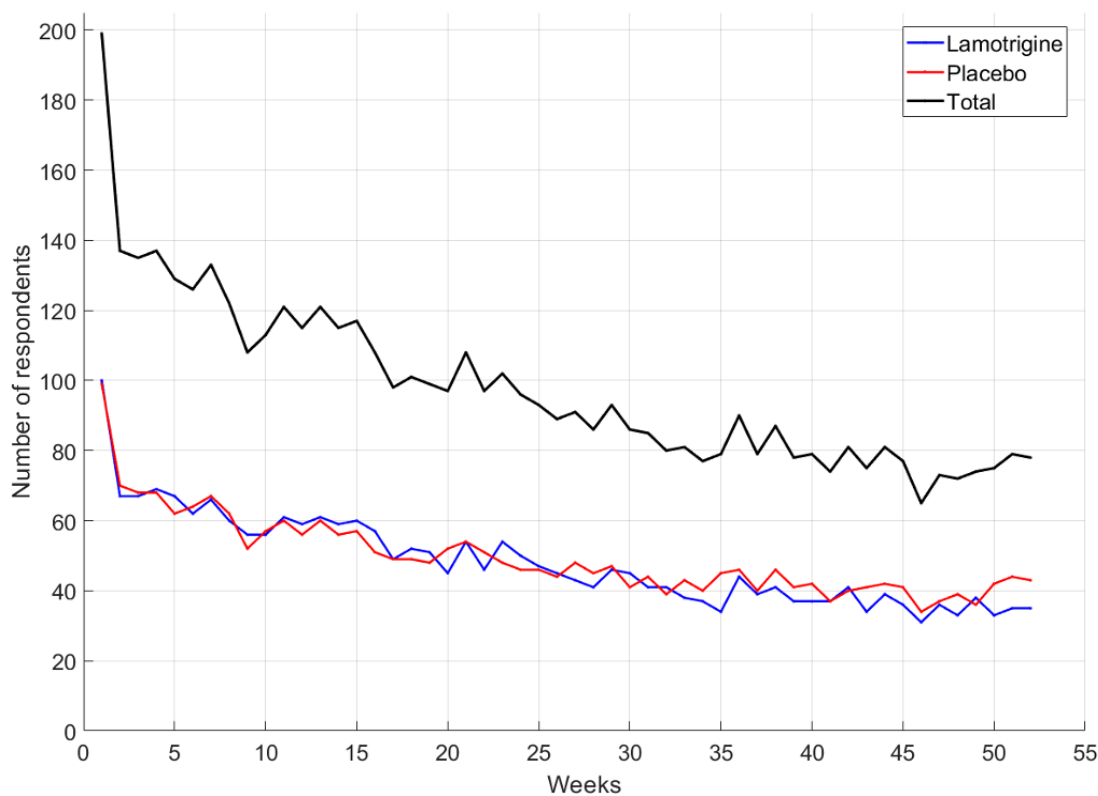

Initially, there were 202 registered participants. After the first 2 weeks, there was a large drop in the number of people who continued to report; however, there were almost equal numbers of subjects in both the lamotrigine and placebo groups. A more detailed trial profile is presented elsewhere [8].

There was a 60% overall compliance with the self-reported mood ratings at 12 weeks; however, even though fewer participants submitted ratings for the entire 52 weeks, no between group differences were observed. For this reason, 2 data subsets were explored; 153 participants who submitted mood data for at least 5 weeks and 138 participants who reported for at least 10 weeks. Another challenge was that the participants did not submit mood ratings at regular time intervals; therefore, the target frequency of one report per week was not always achieved because patients were able to submit scores at any time during the week, which resulted in unevenly sampled data. As a result, extra care was required when using the general time series techniques that had been originally developed for uniformly sampled data.

**Statistical Analysis**

Because the overall goal was to build a binary classifier that could differentiate the patients taking the lamotrigine from those taking the placebo, the first exploratory step was to study the different statistical metrics, called features, that were calculated using the dataset with the aim of identifying the "good" features to feed into the classifier, that is, those features that had sufficient explanatory power to facilitate the classification task. Therefore, how each feature contributed to the classification accuracy when used individually was also investigated.

A common approach when evaluating the explanatory power of a feature has been to assess its associated probability distribution. Kernel smoothing density estimation [12,13] is a nonparametric technique that can estimate the probability distribution of a random variable based on a small data sample. A Gaussian kernel is commonly used with its kernel function being the standard normal distribution because this method is smoother than a histogram when estimating probability distributions.

A receiver operating characteristic (ROC) curve is a graphical plot that shows the classification threshold variations of a binary classifier. Given a labeled dataset, a binary classifier is able to produce the following 4 results from the comparisons of the predicted class to the original labels: true positives, false positives, true negatives, and false negatives. Because the classification threshold is varied, the ROC curve plots the true positive rate versus the false positive rate with the area under the (ROC) curve providing the single metric for the evaluation of the classification model; the larger the area under the curve, the greater the possibility of realizing a high true positive rate and a low false positive rate.

In practice, using one single variable or using all available variables for the classification may not result in an optimal classifier. Various feature selection techniques can be used to select the best model from among the set of available variables, which involves selecting those variables that are representative of the data variability, while improving the classification accuracy. Two types of techniques for feature selection were considered. On one hand, lasso, elasticnet and ridge regression [14] were selected to optimize the deviance of the model. The deviance is calculated based on $L_1$ and $L_2$ norms of the error obtained on a cross-validation set. With these techniques, the best performing features were added to the model one by one up to the point at which the cross-validation error did not reduce. On the other hand, sequential feature selection techniques work in a similar manner but use the misclassification rate as an error metric. We studied both techniques for feature selection to test the robustness of our results.

The different variables computed from the original dataset were investigated as candidate features to feed into the classifiers. The raw dataset contained different subject attributes, such as demographic information (age and gender) and bipolar type, and reported QIDS-SR16 and ASRM values. The age, gender, and bipolar type were kept as they were, and a range of statistics was computed from the QIDS-SR16 values. The ASRM values were not investigated further as these did not indicate any valuable information in the early exploratory stage. Further, because the lamotrigine was being used to treat bipolar depression, the QIDS-SR16 values offered greater information. The QIDS-SR16 mean and SD were the initial candidate features, that is, if the drug worked, the mean QIDS-SR16 value would be lower for the subjects taking the lamotrigine than for those taking the placebo. The coefficient of variation, which is defined as the ratio of SD to the mean, was also used as a separate feature. Other simple statistics considered were the skewness and kurtosis in each subject's QIDS-SR16 time series, which summarized the QIDS-SR16 data distribution but did not reflect the temporal dynamics.

In addition to these basic statistics, other more complicated features were computed from the QIDS-SR16 time series data. The Lomb periodogram [15,16], also known as least-squares spectral analysis, was seen to be an appropriate technique for estimating the power spectral density of the unevenly sampled time series. The Lomb periodogram can be used to estimate the power for a wide range of frequencies. A frequency threshold of 0.2 was arbitrarily defined and 2 features computed, the sum of the low frequency ($f \leq 0.2$) power and the sum of the high frequency ($f > 0.2$) power. The ratio between the amount of power at these high and low frequencies was also calculated as an additional feature.

Detrended fluctuation analysis (DFA) [17] is another commonly used technique for analyzing biomedical signals. DFA basically measures the statistical self-affinity of a signal. McSharry and Malamud give details of the DFA algorithm and related methods in [18]. Essentially, DFA scaling exponent alpha measures the roughness of a time series; for example, white noise, which fluctuates a lot has alpha of 0.5; for pink noise alpha is 1, and a random walk has alpha of 1.5. The DFA exponent alpha was computed for each individual subject and used as another feature in the classification model.

To maximize the classification performance, a number of linear and nonlinear classifiers were also investigated. A linear classifier is a classification algorithm, the objective function of which is a function of a linear combination of features. A binary linear classifier has a linear decision boundary, and a nonlinear classifier has a nonlinear decision boundary. Detailed algorithmic procedures for the classifiers used here are not given and interested readers can consult the relevant references. The linear classifiers investigated were logistic regression [19], linear discriminant analysis [20], and linear support vector machines [21], and the nonlinear classifiers investigated were quadratic discriminant analysis, Gaussian kernel support vector machines, and K-nearest neighbors [22].

In short, the classification performances were evaluated using the different linear and nonlinear methods for each of the individual features and the subset of features obtained in the feature selection step. However, when using complex nonlinear models, it is relatively easy to overfit the data; therefore, to avoid this problem, the evaluation metric was based on the classification accuracy obtained from the out-of-sample data for which 10-fold cross-validation was used to evaluate the out-of-sample performances. The construction, evaluation, and comparison of the many different models for the several features involved an exhaustive search and comparison of the performance of the quantitative classifiers. In reality, because there was no perfect model, the more these different models agreed with each other, the more confidence there was in the results. Parsimonious models are attractive not only because the risk of overfitting is reduced but also because the simpler the model, the easier it is to interpret the results and improve understanding. Data from one, two, and three quarters of the trial period were also considered so that the performance of the classifier could be monitored over time to assess the optimal trial durations for the lamotrigine and placebo comparison.

## Results

The initial data exploration analyzed the trends in the QIDS-SR16 time series. Figure 2 shows the comparative trend analysis for the 2 patient groups, those taking the lamotrigine and those taking the placebo. The results clearly show that there was a more pronounced decreasing trend in the participants taking the lamotrigine. This simple analysis showed that although both groups improved, the rate of mprovement was higher throughout the trial for those taking lamotrigine.

The probability distributions for the reported QIDS-SR16 and ASRM values were also investigated. The kernel density estimates shown in Figure 3 revealed that there was a clear difference between the QIDS-SR16 values reported by the 2 groups. The expanding window average for the QIDS-SR16 and ASRM values are also shown in the same figure. The mean value of the data available up to week n was computed and plotted against the time (number of weeks), which found that the ASRM data were unable to offer any predictive information for the classification of the 2 groups.

The results in Figure 3 show that the overall differences in the QIDS-SR16 scores between the 2 groups was about −2.2 at the end of the 52-week period. Two-tailed $t$ test on the mean QIDS-SR16 scores of each participant were performed for different 4 week periods, and the results are summarized in Table 1. The data were randomly shuffled and the 2 groups randomly selected regardless of whether the samples belonged to the placebo or lamotrigine groups. The same process was repeated to measure the average distribution differences in the QIDS-SR16 values for any 2 groups. The results again indicated that the decrease of 2.2 was statistically significant ($P<.001$).

After designing many features to feed into the classifier, feature selection was conducted using lasso, elasticnet, ridge regression, and the sequential feature selection techniques. All these methods agreed on the choice of only 2 variables, the scaling exponent alpha from DFA and the coefficient of variation ($\sigma/\mu$). The performance evaluations for these 2 features are summarized in Figure 4. The results suggested that the lamotrigine group had a lower DFA exponent alpha, which corresponded to a time series with greater fluctuations, such as in the case of white noise. It appeared that the lamotrigine group had a higher coefficient of variation ($\sigma/\mu$), which corresponded to a lower mean and a higher SD for the QIDS-SR16 values.

These 2 selected features produced easily understood information. The coefficient of variation ($\sigma/\mu$) reflected the shape of the QIDS-SR16 distribution and was not affected by the weekly fluctuations, and the DFA exponent alpha quantified the nature of the temporal weekly fluctuations and was not affected by the distribution. Therefore, intuitively, the coefficient of variation ($\sigma/\mu$) was seen as a standardized measure of the dispersion and the DFA exponent alpha as a measure of the temporal stability.

**Figure 2.** Trend analysis: The linear regression lines for the Quick Inventory of Depressive Symptomatology—self report version 16 (QIDS-SR16) data were plotted.

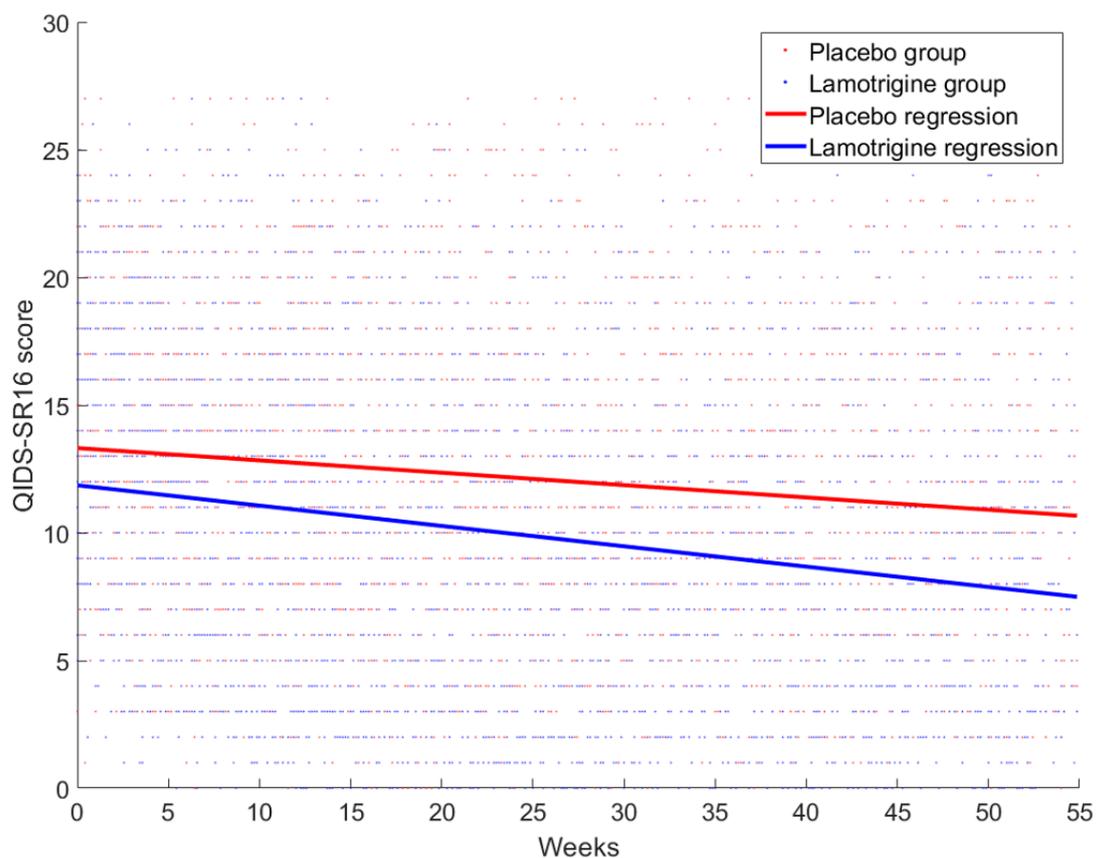

**Figure 3.** Upper half: cumulative distribution function (CDF) density plot for the Quick Inventory of Depressive Symptomatology and Altman self-rating scale value estimates. Lower half: plot for the expanding window average for the Quick Inventory of Depressive Symptomatology—self report version 16 (QIDS-SR16) and Altman self-rating scale (ASRM).

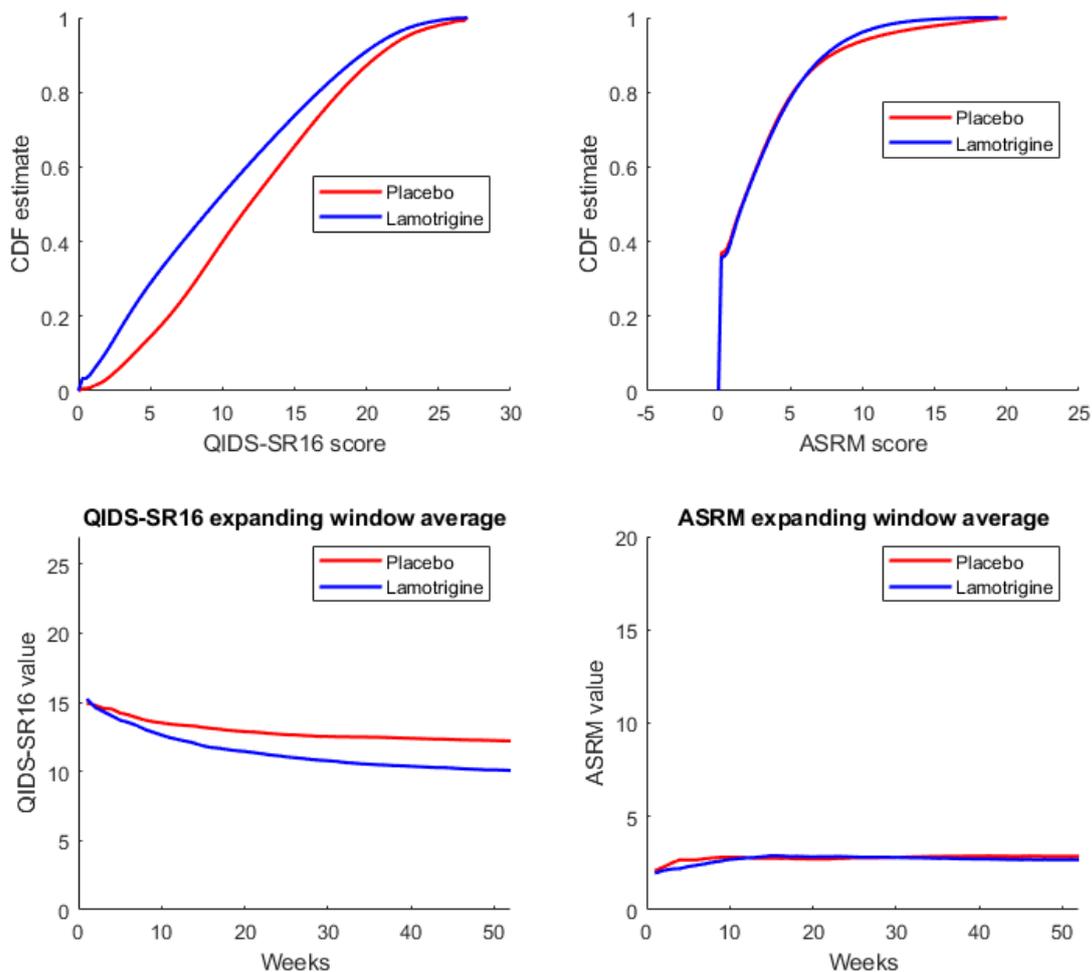

**Table 1.** Results from 2 sample *t* tests on the Quick Inventory of Depressive Symptomatology—self report version 16 (QIDS-SR16) scores for different 4 week periods.

| Four-week period | Sample size | | QIDS-SR16 difference | *P* value | 95% CI |
|---|---|---|---|---|---|
| | Lamotrigine | Placebo | | | |
| 1-12 | 101 | 101 | −0.33 | .63 | −1.02 to 1.69 |
| 10-14 | 81 | 77 | −1.63 | .09 | −0.24 to 3.50 |
| 20-24 | 62 | 64 | −1.52 | .18 | −0.73 to 3.77 |
| 30-34 | 52 | 55 | −3.03 | .007 | 0.83 to 5.23 |
| 40-44 | 50 | 49 | −2.09 | .08 | −0.22 to 4.40 |
| 48-52 | 54 | 48 | −2 | .09 | −0.34 to 4.35 |

**Figure 4.** Cumulative distribution function (CDF) density estimates and receiver operating characteristic (ROC) curves for the selected variables: detrended fluctuation analysis exponent alpha (DFAα) and coefficient of variation (σ/μ).

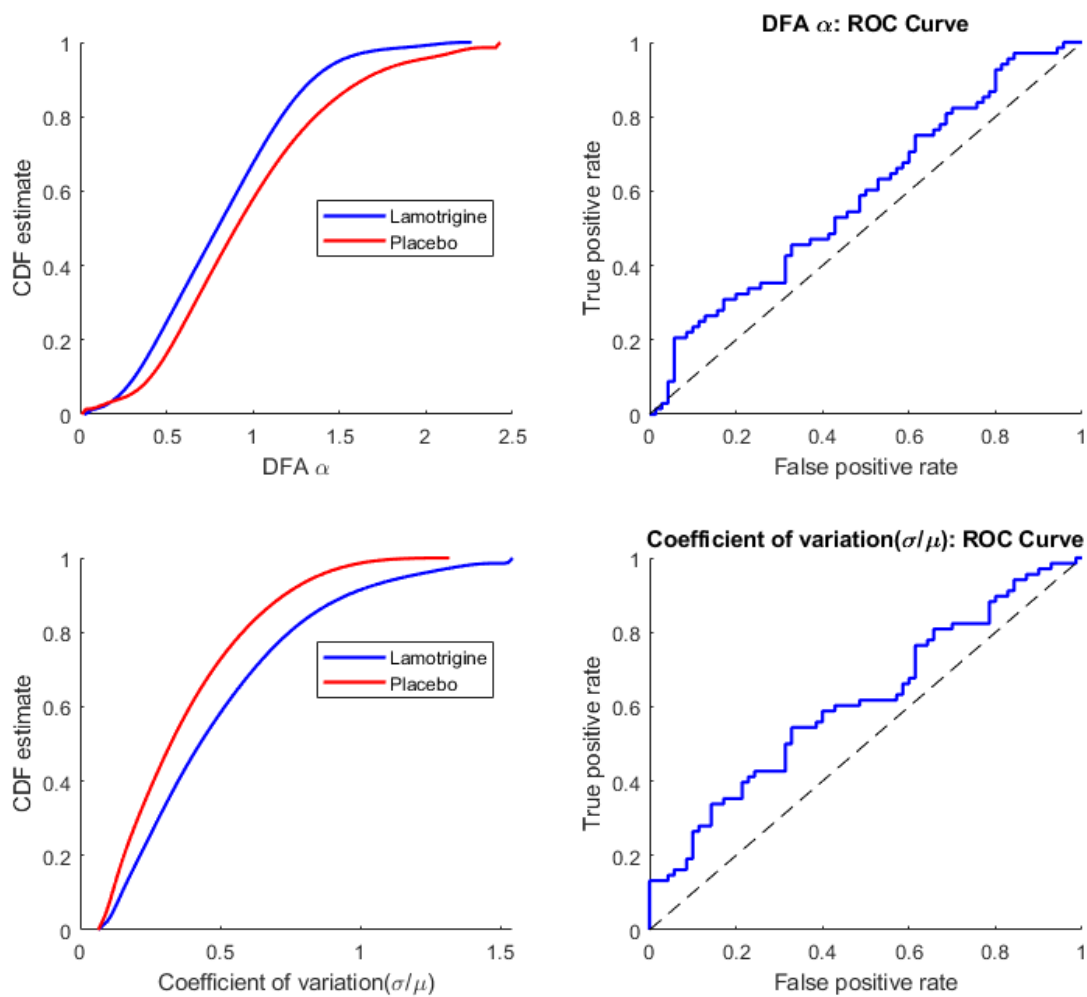

Because the chosen model only had 2 variables, it was easy to visualize the classification decision boundary. Figure 5 is a plot of the decision boundary in a two-dimensional space for the coefficient of variation (σ/μ) versus the DFA exponent alpha. The blue area corresponds to the classification of the lamotrigine group (blue dots), whereas the pink area depicts the classification of the placebo group (red crosses). The results suggested that the lamotrigine group had a higher coefficient of variation and a lower DFA exponent. The decision boundary plot emphasizes the same results that were suggested by the density plots and the ROC curves in Figure 4.

Given the linear separation of the dataset, it was also possible to visualize the time series for the participants lying in the 4 decision area corners; Figure 6 is a plot of these time series. The time series on the left (small alpha) showed greater instability whereas those on the right (large alpha) were smoother and more stable. The time series on the top (large σ/μ) showed less depression on average than those on the bottom (small coefficient of variation [σ/μ]). The time series in the top left corner corresponded to a typical participant in the lamotrigine group (large coefficient of variation [σ/μ] and small alpha), whereas the time series on the bottom right showed a typical participant in the placebo group (small coefficient of variation [σ/μ] and large alpha).

A binary classification analysis was conducted, for which a variety of linear and nonlinear models were applied to the individual features, all features, and to the 2 selected features (coefficient of variation and DFA exponent alpha). The results are summarized in Multimedia Appendix 1. These results suggested that the cross-validation classification accuracy was 62%. There was little evidence of any benefit to be gained from using the nonlinear classification models; therefore, the study was continued using logistic regression on the 2 selected features.

As mentioned, participant compliance was important and was found to yield time series of varying durations. A clinician who was using the classifier to test whether lamotrigine was better than the placebo would need to know how many weeks of data were required before a definitive decision could be made. For this reason, the performance of the logistic regression model was evaluated against the amount of data reported by participants. For this evaluation, the participants were selected if they had provided at least 10 QIDS-SR16 responses for trial durations of 13 to 52 weeks. Figure 7 shows the classification accuracy and SE versus the trial duration. The classification accuracy was plotted using available data after each week (from the 13th to the 52nd weeks). The maximum of the fractions for the lamotrigine and placebo participants served as the "no information" benchmark for the classifier. In addition, for most periods from 20 weeks onwards, the classifier was outperforming the benchmark. For trials ranging from 44 to 52 weeks, the classification accuracy was greater than the benchmark and was statistically significant.

**Figure 5.** Classification decision boundary using logistic regression. BN: Brownian noise; DFAα: detrended fluctuation analysis alpha; N: pink noise; WN: white noise; σ/μ: coefficient of variation.

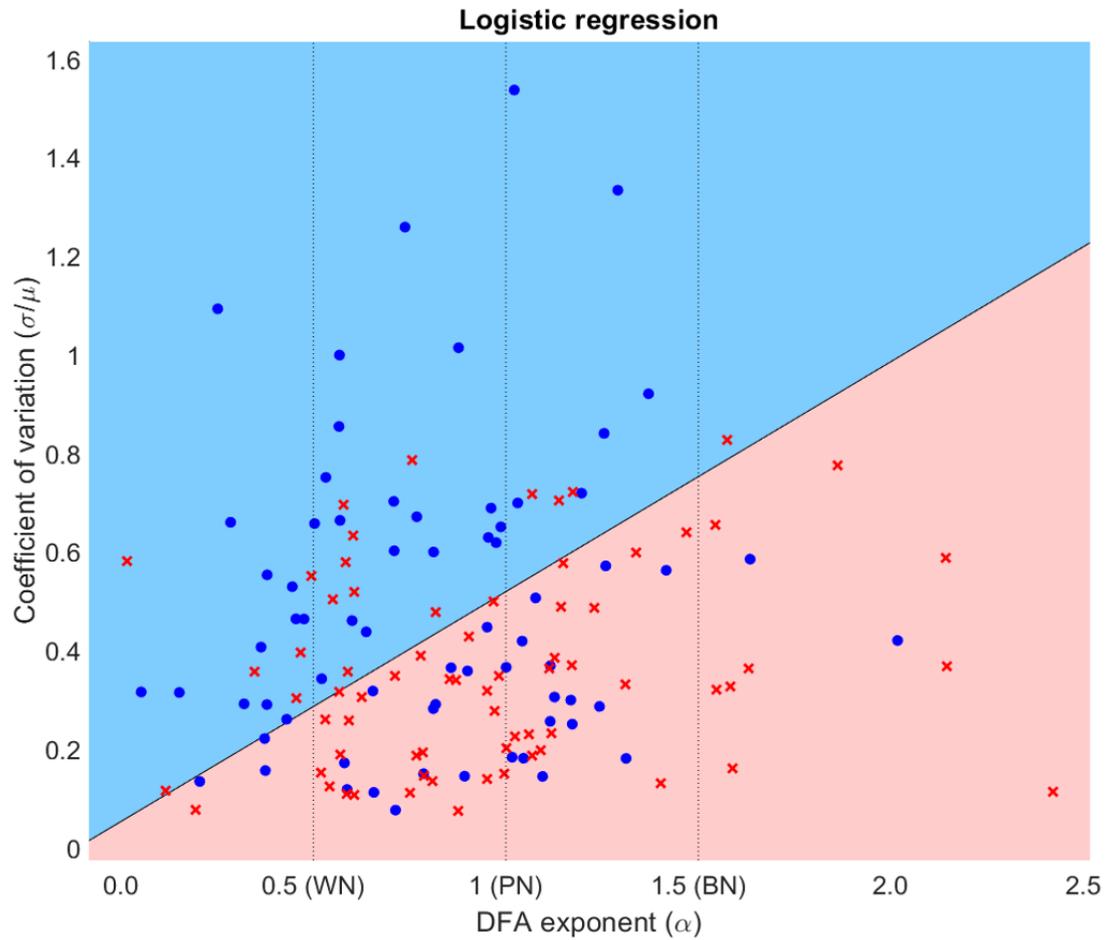

**Figure 6.** The four Quick Inventory of Depressive Symptomatology—self report version 16 (QIDS-SR16) time series for the subjects lying near the 4 decision boundary plot corners.

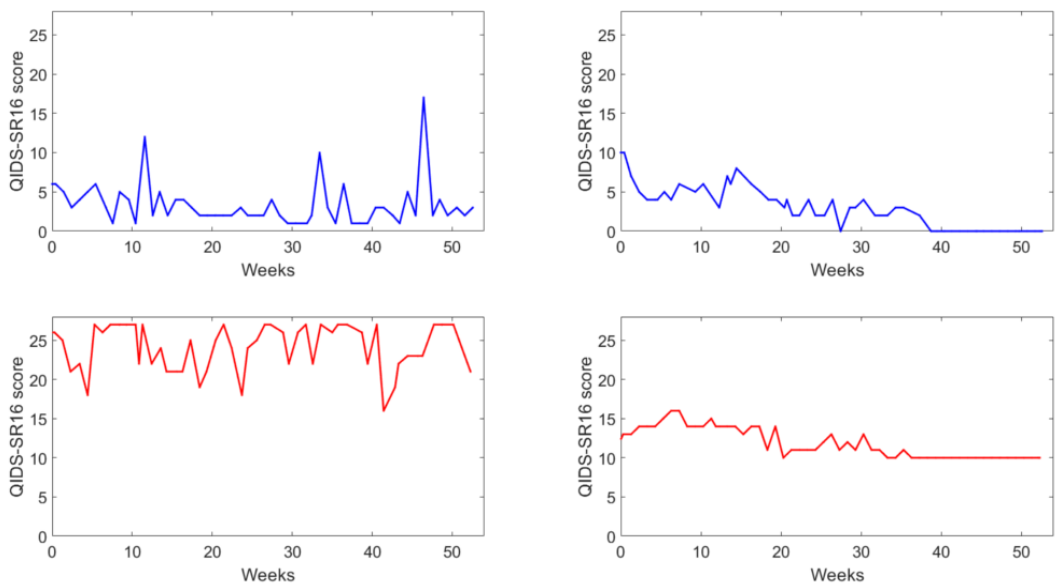

**Figure 7.** Data requirements for individual subjects and classification accuracy for data available after 13 weeks.

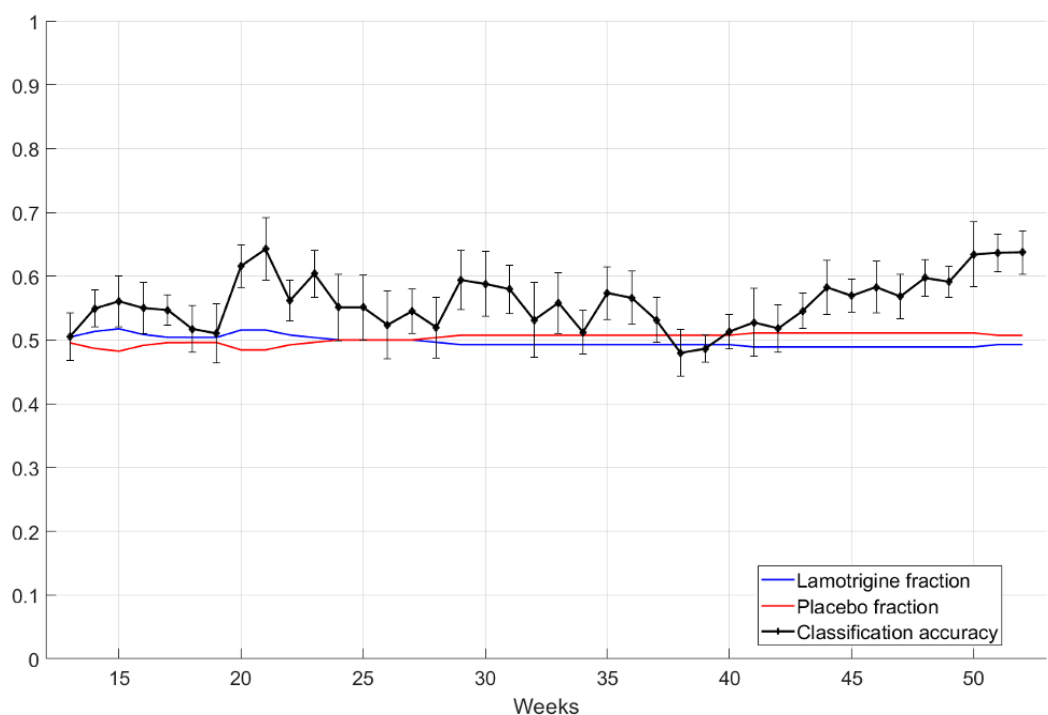

## Discussion

### Methods and Their Limitations

The results from the classification model confirmed the original findings of the CEQUEL study that the addition of lamotrigine to quetiapine for the treatment of severe bipolar depression decreased depressive symptoms compared with a placebo. Although the original CEQUEL data analysis had relied on linear regression model fitting, the classification model used in this study not only examined the data from a different perspective but also provided robust explanatory power because it allowed for an estimation of the extent to which the 2 groups of observations (ie, treatment allocation: active lamotrigine vs placebo) could be distinguished based only on the QIDS-SR16 scores. Therefore, a model was constructed without the need for any prior information about the clinical interventions during the treatment period. Using out-of-sample classification accuracy and simpler models allowed us to avoid overfitting that is problematic when employing complicated machine learning models on shallow datasets. Another advantage of the machine learning models was the ability to test multiple features computed from the raw data and select only those features that had sufficient explanatory power.

The use of DFA on the QIDS-SR16 time series allowed us to examine the temporal stability of the treatment, an aspect that was not considered in the original study. Future data analyses could attempt to explain why the QIDS-SR16 scores of the participants taking the active lamotrigine were generally temporally unstable. The effect of other clinical interventions and concurrent treatments during the study period could also be considered. Finally, it was demonstrated that machine learning techniques could be generally used in clinical trials to provide greater insights into what the data represent beyond classical statistical analyses, especially when there are large, complex datasets available. One drawback of using machine learning techniques, however, is that the analyst must deal with the bias-variance tradeoff. Another disadvantage is that some powerful machine learning models require very large datasets to be able to generalize well.

### Conclusions

This study confirmed that the use of lamotrigine decreases depressive symptoms in bipolar patients. The selected classification features suggested that lamotrigine increased the coefficient of variation (achieved by increasing SD or decreasing the mean of the QIDS-SR16 time series). It was also found that patients taking lamotrigine tended to have rougher time series, which was indicative of a greater temporal instability in the time series. The 2 features, the coefficient of variation and DFA exponent, implied that a two-dimensional visualization diagram and linear decision boundary can be constructed to better understand bipolar disorder and the ways that the participants are affected by lamotrigine. The statistical significance of the classification was evaluated, from which it was determined that a trial of at least 44 weeks was required to distinguish between lamotrigine and the placebo. It would be useful to conduct additional studies to obtain a larger cohort of compliant participants. The selected features provided a deeper understanding of the temporal dynamics of subjects experiencing bipolar disorder and offered the potential for the better monitoring of symptoms over time.

## Conflicts of Interest

None declared.

## Multimedia Appendix 1

Comparison of models: Logistic regression (LR), Linear Discriminant Analysis (LDA), Quadratic Discriminant Analysis (QDA), Linear Support Vector Machine (LSVM), Gaussian Kernel SVM (GSVM) and K-Nearest Neighbors (KNN). We show both the in-sample and out-of-sample classification accuracy.

[PDF File (Adobe PDF File), 47KB-Multimedia Appendix 1]

## References

1. Anderson IM, Haddad PM, Scott J. Bipolar disorder. BMJ 2012 Dec 27;345(dec27 3):e8508-e8508. [doi: 10.1136/bmj.e8508]
2. Merikangas KR, Akiskal HS, Angst J, Greenberg PE, Hirschfeld RMA, Petukhova M, et al. Lifetime and 12-month prevalence of bipolar spectrum disorder in the National Comorbidity Survey replication. Arch Gen Psychiatry 2007 May;64(5):543-552 [FREE Full text] [doi: 10.1001/archpsyc.64.5.543] [Medline: 17485606]
3. Schmitt A, Malchow B, Hasan A, Falkai P. The impact of environmental factors in severe psychiatric disorders. Front Neurosci 2014;8:19 [FREE Full text] [doi: 10.3389/fnins.2014.00019] [Medline: 24574956]
4. Saunders KEA, Goodwin GM. The course of bipolar disorder. Adv Psychiatr Treat 2018 Jan 02;16(5):318-328. [doi: 10.1192/apt.bp.107.004903]
5. Sidor MM, MacQueen GM. An update on antidepressant use in bipolar depression. Curr Psychiatry Rep 2012 Dec;14(6):696-704. [doi: 10.1007/s11920-012-0323-6] [Medline: 23065437]
6. Geddes JR, Calabrese JR, Goodwin GM. Lamotrigine for treatment of bipolar depression: independent meta-analysis and meta-regression of individual patient data from five randomised trials. Br J Psychiatry 2009 Jan;194(1):4-9. [doi: 10.1192/bjp.bp.107.048504] [Medline: 19118318]
7. Brodie MJ. Lamotrigine. Lancet 1992 Jun 06;339(8806):1397-1400. [Medline: 1350815]
8. Geddes J, Gardiner A, Rendell J, Voysey M, Tunbridge E, Hinds C, et al. Comparative evaluation of quetiapine plus lamotrigine combination versus quetiapine monotherapy (and folic acid versus placebo) in bipolar depression (CEQUEL): a 2 × 2 factorial randomised trial. The Lancet Psychiatry 2016 Jan;3(1):31-39. [doi: 10.1016/S2215-0366(15)00450-2]
9. Rush A, Trivedi M, Ibrahim H, Carmody T, Arnow B, Klein D, et al. The 16-Item Quick Inventory of Depressive Symptomatology (QIDS), clinician rating (QIDS-C), and self-report (QIDS-SR): a psychometric evaluation in patients with chronic major depression. Biol Psychiatry 2003 Sep 01;54(5):573-583. [Medline: 12946886]
10. Altman EG, Hedeker D, Peterson JL, Davis JM. The Altman Self-Rating Mania Scale. Biol Psychiatry 1997 Nov 15;42(10):948-955. [doi: 10.1016/S0006-3223(96)00548-3] [Medline: 9359982]
11. Bopp J, Miklowitz D, Goodwin G, Stevens W, Rendell J, Geddes J. The longitudinal course of bipolar disorder as revealed through weekly text messaging: a feasibility study. Bipolar Disord 2010 May;12(3):327-334 [FREE Full text] [doi: 10.1111/j.1399-5618.2010.00807.x] [Medline: 20565440]
12. Rosenblatt M. Remarks on Some Nonparametric Estimates of a Density Function. The Annals of Mathematical Statistics 1956 Sep;27(3):832-837 [FREE Full text]
13. Parzen E. On Estimation of a Probability Density Function and Mode. The Annals of Mathematical Statistics 1962 Sep;33(3):1065-1076 [FREE Full text]
14. Tibshirani R. Regression Shrinkage and Selection via the Lasso. Journal of the Royal Statistical Society. Series B (Methodological) 1996;58(1):267-288 [FREE Full text]
15. Lomb N. Least-squares frequency analysis of unequally spaced data. Astrophysics and Space Science 1976 Feb;39(2):447-462 [FREE Full text] [doi: 10.1007/BF00648343]
16. Scargle JD. Studies in astronomical time series analysis. II - Statistical aspects of spectral analysis of unevenly spaced data. Astrophysical Journal 1982 Sep 15;263:835-853 [FREE Full text]
17. Peng CK, Buldyrev SV, Havlin S, Simons M, Stanley HE, Goldberger AL. Mosaic organization of DNA nucleotides. Phys Rev E Stat Phys Plasmas Fluids Relat Interdiscip Topics 1994 Feb;49(2):1685-1689. [Medline: 9961383]
18. McSharry PE, Malamud BD. Quantifying self-similarity in cardiac inter-beat interval time series. : IEEE; 2005 Sep Presented at: Computers in Cardiology 2005; 25-28 September 2005; Lyon, France URL:https://ieeexplore.ieee.org/abstract/document/1588136/ [doi: 10.1109/CIC.2005.1588136]
19. Cox DR. The Regression Analysis of Binary Sequences. Journal of the Royal Statistical Society. Series B (Methodological) 1958;20(2):215-242 [FREE Full text]
20. McLachlan G. Discriminant analysis and statistical pattern recognition. Hoboken, New Jersey, United States of America: John Wiley & Sons; Aug 04, 2004.
21. Cortes C, Vapnik V. Support-vector networks. Machine Learning 1995;20(3):273-297. [doi: 10.1007/BF00994018]
22. Altman NS. An Introduction to Kernel and Nearest-Neighbor Nonparametric Regression. The American Statistician 1992;46(3):175-185 [FREE Full text]

## Abbreviations

**ASRM:** Altman self-rating scale
**CEQUEL:** Comparative evaluation of Quetiapine-Lamotrigine
**DFA:** detrended fluctuation analysis
**QIDS-SR16:** Quick Inventory of Depressive Symptomatology–Self-Report version 16
**ROC:** receiver operating characteristic